\documentclass[aps,prl,showpacs,twocolumn,amsmath,amssymb,nofootinbib]{revtex4-1}
\usepackage{graphicx}
\usepackage{dcolumn}
\usepackage{bm}
\usepackage{amsmath}
\usepackage{amsfonts}
\usepackage{amssymb}
\usepackage{multirow}

\usepackage{amsthm}
\usepackage{pinlabel}
\usepackage{natbib}

\usepackage{epstopdf}
\usepackage[abs]{overpic}
\usepackage[dvipsnames]{xcolor}

\usepackage[normalem]{ulem}
\usepackage{ulem}

\usepackage[dvipsnames]{xcolor}

\newcommand{\ph}{^{\phantom{\dagger}}}
\usepackage{braket}

\definecolor{specialgray}{HTML}{505050}
\definecolor{col10K}{HTML}{FFA000}
\definecolor{col300K}{HTML}{924FA4}
\definecolor{colMu}{HTML}{5278BD}
\definecolor{colMuI}{HTML}{924FA4}

\definecolor{specialgray}{HTML}{505050}
\definecolor{col10K}{HTML}{FFA000}
\definecolor{col300K}{HTML}{924FA4}
\definecolor{colMu}{HTML}{5278BD}
\definecolor{colMuI}{HTML}{924FA4}
\definecolor{newred}{HTML}{D53E4F}
\definecolor{newblue}{HTML}{5278BD}
\definecolor{newcyan}{HTML}{1EA0A0}
\definecolor{newgreen}{HTML}{5CB14E}
\definecolor{newpurple}{HTML}{924FA4}
\definecolor{newyellow}{HTML}{D1C72E}
\definecolor{neworange}{HTML}{D6923C}

\begin{document}
\title{Unconventional superconductivity mediated  solely by isotropic electron-phonon interaction}
\author{Fabian Schrodi}\email{fabian.schrodi@physics.uu.se}
\author{Peter M. Oppeneer}\email{peter.oppeneer@physics.uu.se}
\author{Alex Aperis}\email{alex.aperis@physics.uu.se}
\affiliation{Department of Physics and Astronomy, Uppsala University, P.\ O.\ Box 516, SE-75120 Uppsala, Sweden}

%\vskip 0.4cm
\date{\today}

\begin{abstract}
	\noindent 
Unconventional superconductivity is commonly linked to electronic pairing mechanisms, 
since it is believed that
the conventional electron-phonon interaction (EPI) cannot cause sign-changing superconducting gap symmetries. Here, we show that this common understanding needs to be revised
when one considers a more elaborate theory of electron-phonon superconductivity beyond standard approximations. 
We selfconsistently solve the full-bandwidth, anisotropic Eliashberg equations including vertex corrections beyond Migdal's approximation assuming the usual isotropic EPI for cuprate, Fe-based and heavy-fermion superconductors with nested Fermi surfaces.
In case of the high-$T_c$ cuprates
we find a $d$-wave order parameter, as well as a nematic state upon increased
doping. For Fe-based superconductors, we obtain  $s_{\pm}$ gap symmetry, while for heavy-fermion CeCoIn$_5$ we find unconventional $d$-wave pairing. These results provide a proof-of-concept that 
EPI cannot be excluded as a mediator of unconventional and of high-$T_c$ superconductivity.
\end{abstract}

\maketitle

Superconductors with unconventional---that is non-$s$-wave---gap symmetries continue to attract {great}
interest, because of their unique properties \cite{Sigrist1991,Qi2011} and because they are closely linked to the as yet unexplained phenomenon of high-$T_c$ superconductivity \cite{100yrs,Keimer2015}.
%Unconventional superconductors  attract immense scientific interest,  partly due to the unique properties  deriving from their non s-wave Cooper pairing  \cite{Sigrist1991,Qi2011} but most importantly because they are intimately linked to the yet unexplained phenomenon of high-T$_c$ superconductivity \cite{100yrs}. 
Prominent material examples are the {high-$T_c$}
%$d_{x^2-y^2}$ 
cuprates {with $d$-wave} and the iron-based superconductors with predominantly $s_\pm$ {gap symmetry}
\cite{Tsuei2000,Stewart2011}. Members of these families share several common features like e.g., quasi two-dimensional electronic bandstructures with good nesting properties, proximity to magnetic ordering and strongly coupled superconductivity beyond the Bardeen-Cooper-Schrieffer (BCS) picture \cite{Uemura1989}. Resonances observed in inelastic neutron scattering {experiments} {suggest} the presence of spin fluctuations {in the pairing mechanism}  \cite{Tsuei2000,Dai2015, Scalapino2012} but large isotope effects pointing at the involvement of phonons have also been measured \cite{Gweon2004,Iwasawa2008,Liu2009,Khasanov2010}. These phonons have characteristic energies of similar order of magnitude {as those} of the relevant spin fluctuations  \cite{Lanzara2001}. 
In this setting, the key for 
{pinpointing} the driving {pairing} mechanism has been the symmetry of the superconducting gap itself \cite{Tsuei2000, Scalapino2012}.

%For decades, 
An unconventional gap {is commonly considered as the signature}
of a repulsive, solely electronic mechanism that can be pairing if the electrons allow a sign-change in their wavefunction. The most prominent example is the antiferromagnetic spin-fluctuations mechanism where the wavevector of the dominant magnon excitations, which is associated with the nesting properties of the underlying electronic bandstructure of the material, matches the sign-change of the superconducting gap \cite{Scalapino1986}. Many superconductors comply with this picture, including cuprates, iron-based and heavy-fermion materials \cite{Scalapino2012}.

%Contrasting the above, 
{Conversely,} in the original theory of BCS \cite{Bardeen1957} and its refinement by Eliashberg to include retardation effects \cite{Eliashberg1960}, phonons mediate an effective attractive interaction between electrons which gives rise to a sign-preserving $s$-wave gap. 
%There have been few 
Attempts to reconcile the phonon mechanism with unconventional gap symmetries do exist. For example, it has been demonstrated that a small-q-peaked electron-phonon interaction (EPI) can give rise to the unconventional gap of the cuprates and the Fe-pnictides in the presence of enhanced Coulomb repulsion between electrons \cite{Varelogiannis1998,Aperis2011}. 
Despite recent observations of such small-q EPI in monolayer FeSe/SrTiO$_3$ \cite{Lee2014}, its existence in bulk {superconductors} 
is yet to be confirmed. 

Notably,  
 Eliashberg's theory for boson-mediated  superconductivity rests on the premise of adiabaticity, i.e.\ the energy scale of electrons, $\epsilon_F$, is much larger than that of the relevant bosons, $\Omega$. In the case of phonons, Migdal showed that when the so-called nonadiabatic ratio $\alpha=\Omega/\epsilon_F$ is small ($\sim 10^{-2}$), vertex corrections to the electron self-energy due to the EPI are negligible \cite{Migdal1958}. 
 {However,} for many unconventional superconductors $\alpha\sim 0.1-0.3$ \cite{Uemura1989,Benedetti1994} 
 and thus vertex corrections can not be safely neglected. 
 {Including these} requires solving the selfconsistent Nambu-Dyson equation 
 {with} crossing self-energy diagrams while retaining full frequency and momentum dependence, a task that has only recently been achieved \cite{Schrodi2020}. Due to the numerical 
 {complexity} of this problem, prior attempts adopted {serious} simplifications   \cite{Grimaldi1995,Botti2002,Miller1998,Hague2003}. 
 %{\blue Only one such prior work included momentum variations of the gap {\green via a cluster dynamical mean-field theory approximation} {\red needed?-later!} and showed that a $d_{x^2-y^2}$ gap solution is possible for a purely phononic Holstein model \cite{Hague2003}.}

%\textbf{Methodology.} 
Here, we %revisit the longstanding problem of 
{address the long-standing issue of} phonon-mediated unconventional superconductivity by {performing}
direct numerical solutions of the full-bandwidth anisotropic Eliashberg equations including the first vertex corrections to the self-energy beyond Migdal's approximation using material specific input for three archetypal unconventional superconducting systems: cuprates, iron pnictides and the heavy fermion CeCoIn$_5$ {that are known to exhibit  quasi two-dimensional electronic structures as well as nested Fermi surfaces (FSs).}
In all our calculations, we assume an
optical phonon mode that couples to electrons via a purely isotropic EPI which is the only interaction in the system. 

%%%Surprisingly, 
%We find that such vertex corrected {conventional} EPI leads not only to a pure $d_{x^2-y^2}$ but also to a nematic superconducting gap in cuprates, an $s_{\pm}$ gap in iron pnictides and a higher-harmonic $d_{x^2-y^2}$ gap in CeCoIn$_5$.  {\blue We show that} this phonon-mediated unconventional superconductivity relies on Fermi surface (FS) nesting and nonadiabaticity. The latter give{\red s} rise to a repulsive EPI which is anti-pairing in the $s$-wave channel \cite{Schrodi2020} while the former makes this interaction peak near the nesting  wavevector(s) and thus favorable for unconventional pairing in a similar fashion as purely electronic mechanisms like spin fluctuations. 
%Our results {\blue show that the conventional EPI can mediate unconventional superconductivity and highlight}
%%%prove that, in contrast to the  broadly accepted picture, {\em the conventional EPI is perfectly compatible with 
%%%unconventional superconductivity}. Given the unambiguous existence of phonons in all crystalline solids, these findings 
%%%emphasize 
%the possibility that electron-phonon coupling plays a substantial role in any superconductor with unconventional gap symmetry including the high-T$_c$s.

We {start with describing our} 
model where electrons 
forming $n$ energy bands with dispersions $\xi_{{\bf k},n}$ couple to Einstein phonons of frequency $\Omega$ via an isotropic and band-independent EPI with coupling strength $g_{\mathbf{q}}=g_0$, where $\mathbf{q}=\mathbf{k}-\mathbf{k}'$. We consider the following  Nambu form of the vertex-corrected electron self-energy for such system,
\begin{eqnarray}\label{selfen}
\hat{\Sigma}_{{\bf k},m}=T\sum_{{\bf k}',n,m'}g^2_0 D_{m-m'}\hat{\rho}_3 \hat{G}_{{\bf k}',n,m'}\hat{\rho}_3 \, \hat{\Gamma}\ph_{{\bf q},m,m'} \, ,
\end{eqnarray}
with Matsubara frequencies $\omega_m=\pi T(2m+1)$, $m\in\mathbb{Z}$, at temperature $T$. For brevity we employ here the notation $f(\mathbf{k},i\omega_m)=f_{\mathbf{k},m}$ for any function $f$. The vertex renormalization function (hereafter denoted just vertex function) is given by,  
\begin{eqnarray}\label{vertex}
\hat{\Gamma}\ph_{{\bf q},m,m'} = 1 + T\sum_{m''}g^2_0 D_{m'-m''}\hat{\Lambda}\ph_{{\bf q},m'',m''-m'+m} \,,
\end{eqnarray}
with
\begin{eqnarray}\label{suscl}
\hat{\Lambda}\ph_{{\bf q},m'',m''+l}=  \!\!\sum_{{\bf k}'',n',n''}  \!\! \hat{G}_{{\bf k}'',n',m''}  \hat{\rho}_3\hat{G}_{{\bf k}''+{\bf q},n'',m''+l}\hat{\rho}_3 \end{eqnarray}
and $l=m-m'$. Eq.\,(\ref{selfen}) includes an infinite series of Feynman diagrams that constitute the lowest order Migdal self-energy and 
an infinite series of crossing diagrams that {constitute}
the first vertex correction beyond Migdal's approximation \cite{Botti2002,Schrodi2020}. {Further details can be found in the Supplemental Material (SM) \cite{supplement}.} 
In the above, $D_{m-m'}$ is the phonon propagator,
% \cite{supplement}, 
while the electron Green's function in Nambu space with Pauli basis $\hat{\rho}_i$ is given by
\begin{align}
\hat{G}^{-1}_{\mathbf{k},n,m} =& \, i\omega_m Z_{\mathbf{k},m}\hat{\rho}_0 - \phi_{\mathbf{k},m}\hat{\rho}_1 - [\xi_{\mathbf{k},n} +\zeta_{\mathbf{k},m}]\hat{\rho}_3 \,. \label{greensfun}
\end{align}
Here $Z_{\mathbf{k},m}$, $\zeta_{\mathbf{k},m}$ and $\phi_{\mathbf{k},m}$ describe, respectively, the mass and chemical potential renormalization, and the superconducting order parameter. From Eqs.\,(\ref{selfen})-(\ref{suscl}) and 
(\ref{greensfun}) we derive a set of vertex-corrected, anisotropic, full bandwidth and multi-band Eliashberg equations for these three functions, which are solved selfconsistently without further approximation\,\cite{Schrodi2020,UppSC,supplement}.
%\cite{supplement}).} 
From the results we obtain the gap function via $\Delta_{\mathbf{k},m}=\phi_{\mathbf{k},m}/Z_{\mathbf{k},m}$, and the gap edge $\Delta(\mathbf{k})\approx\Delta_{\mathbf{k},{m=0}}$.

%{\blue To start with the cuprate superconductors, we adopt for the electron band structure}
%In our calculations 
We model the electron band structure of cuprate superconductors with 
the commonly employed tight-binding model, $\xi_{\mathbf{k}}=\beta_{\mathbf{k}}-\delta_{\mathbf{k}}-\mu$, that includes nearest and next-nearest neighbor hopping terms $\beta_{\mathbf{k}}=t\,[\cos k_x + \cos k_y ]$ and $\delta_{\mathbf{k}}=C-t' \cos k_x \cos k_y$, respectively. Unless noted otherwise we {set} the hopping energies $t$, $t'$, parameter $C$, and chemical potential $\mu$ to $(C,t,t',\mu)=(0,-0.25,0.1,-0.07)$ eV. We further {adopt} $g_0=148\,\mathrm{meV}$ and choose a phonon frequency $\Omega=50\,\mathrm{meV}$ that is motivated from Angle Resolved Photoemission Spectroscopy (ARPES) measurements on different cuprate systems\,\cite{Lanzara2001}. For Fe-based materials the scattering strength and characteristic phonon frequency are chosen as $g_0=130\,\mathrm{meV}$ and $\Omega=17\,\mathrm{meV}$\,\cite{Kordyuk2011,Zbiri2009,Boeri2009}. The respective  electronic energies are described by a two-band model with $\xi^{(\pm)}_{\mathbf{k}}=\beta_{\mathbf{k}} \pm \delta_{\mathbf{k}} - \mu$ and parameters $(C,t,t',\mu)=(1/3,1/6,1/12,0)\,\mathrm{eV}$. In the case of CeCoIn$_5$ we use {the} two-band tight-binding model of Ref.\,\cite{VanDyke2014} {(which neglects several small FS pockets)} and choose $\Omega=5$\,meV \cite{Martinho2007} and $g_0=4$\,meV. Qualitatively similar results as presented here are also found for larger frequencies and different $g_0$ values {(see SM \cite{supplement}).}
%\cite{supplement}). }
%
\begin{figure*}[t!]
	\centering
	\includegraphics[width=\linewidth]{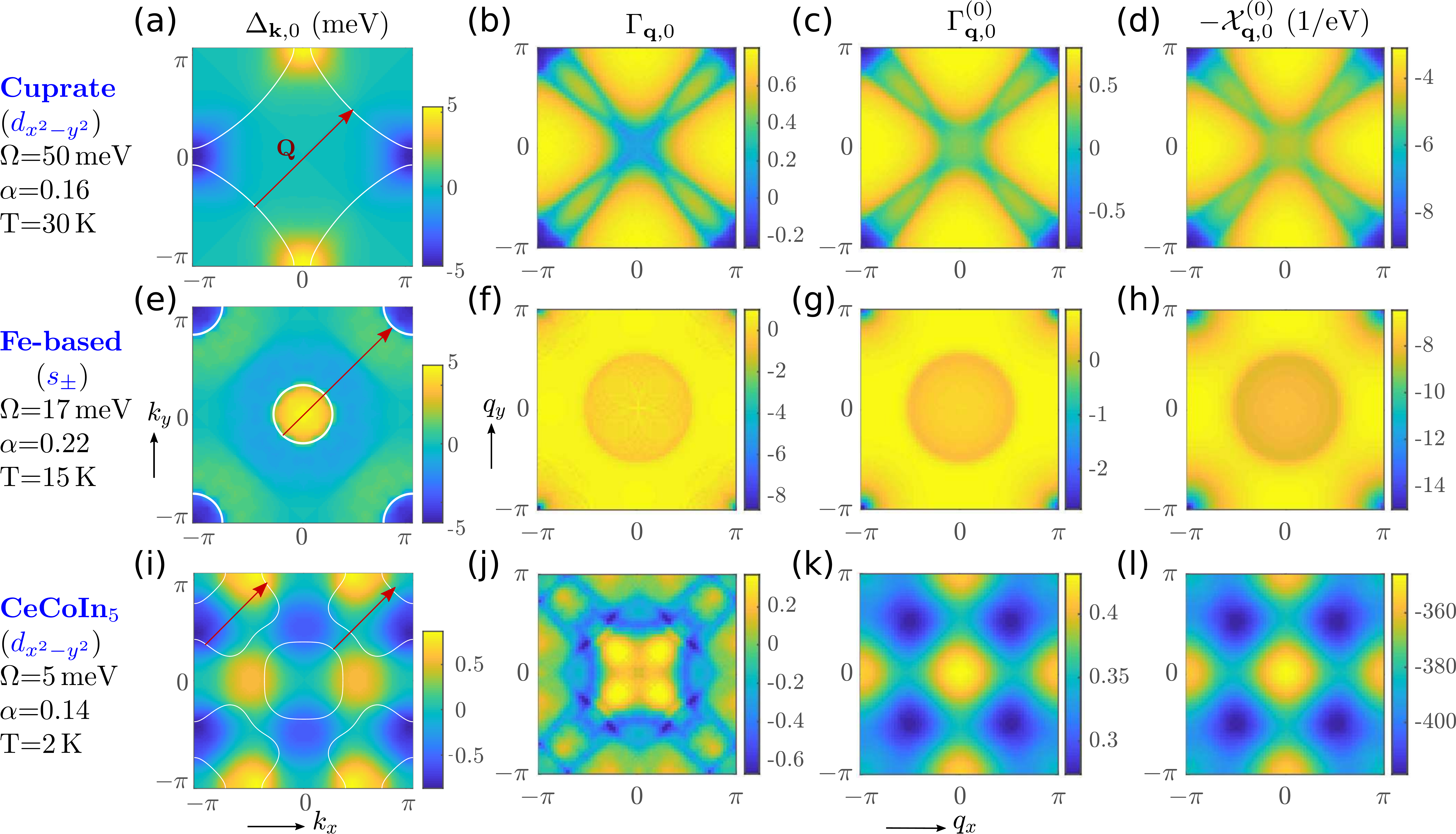}
	\vspace*{-0.6cm}
	\caption{{Selfconsistently calculated unconventional superconductivity.} (a), (e), (i) The unconventional superconducting gap calculated by direct solution of the vertex-corrected Eliashberg equations. The Fermi surface %of the tight-binding model employed 
	is in each case drawn with white lines {in the Brillouin zone}. {Fermi surface nesting is indicated by the red vector $\bf Q$.} (b), (f), (j) The selfconsistently calculated  vertex function, and (c), (g), (k), the calculated bare vertex function for $T>T_c$. (d), (h), (l) The calculated bare charge susceptibility. The first row corresponds to results for cuprates, where a $d_{x^2-y^2}$ gap symmetry is {obtained}. The second row corresponds to Fe-based materials where an $s_\pm$ symmetry is found. The third row corresponds to 
	CeCoIn$_5$ where we {obtain} a higher-harmonic $d_{x^2-y^2}$ gap symmetry. {The used phonon frequency $\Omega$, nonadiabaticity ratio $\alpha$, and temperature $T$ are given in the legend.}}
	\label{symmetry}
\end{figure*}
%

%\textbf{Calculated results.} 
The resulting Fermi surfaces for our cuprate, Fe-based, and CeCoIn$_5$ systems are drawn with white lines in Figs.\,\ref{symmetry}(a),(e), and (i), respectively. The corresponding calculated non-interacting charge susceptibilities,
\begin{eqnarray}
{\cal X}^{(0)}_{{\bf q},0}
=\sum_{{\bf k},n,n'}\frac{n_F(\xi_{{\bf k},n})-n_F(\xi_{{\bf k+q},n'})}{\xi_{{\bf k},n}-\xi_{{\bf k+q},n'}}\ ,
\end{eqnarray}
{with Fermi-Dirac functions $n_F(\cdot)$,}
 are shown in Figs.\,\ref{symmetry}(d), (h), and (l). As expected, these exhibit pronounced peaks at wavevectors ${\bf Q}=(\pi,\pi)$ for cuprate and Fe-based systems, and ${\bf Q}\approx (\pi/2,\pi/2)$ for CeCoIn$_5$. These pronounced peaks evidence directly the well-known good nesting properties of these three quasi-2D systems. {Note that for CeCoIn$_5$ there are several FS parts involved which causes a broadening of the $ (\pi/2,\pi/2)$-peaks.} In spin-fluctuation theories for these systems, such susceptibilities provide the necessary momentum-dependent repulsive interaction that gives rise to unconventional $\Delta ({\bf k})$ symmetries (see \cite{Scalapino2012}). 
 {Here we solve numerically the Eliashberg equations derived from Eqs.\,(\ref{selfen})-({\ref{suscl}}) to determine the gap symmetry.}

The main results of this work are shown in Figs.\,\ref{symmetry}(a), (e), and (i) where we plot our selfconsistently calculated momentum dependence of the superconducting gap for each of the three systems that we consider. Remarkably, as shown in Fig.\,\ref{symmetry}(a), for the cuprate case we find a clear $d_{x^2-y^2}$-gap with a Brillouin-zone (BZ) modulation that is roughly proportional to $\cos{ k_x }-\cos{ k_y }$ and a realistic amplitude ${\mathrm{max}}_{\mathbf{k}}\,\Delta(\mathbf{k})\simeq5.4\,\mathrm{meV}$ at $T=30$\,K. 
{We note that previously it was found that a $d_{x^2-y^2}$ symmetry is possible for a pure phononic Holstein model using 
 a cluster dynamical mean-field theory approximation \cite{Hague2003}.}
For the Fe-based system, shown in Fig.\,\ref{symmetry}(e), we find the $s_\pm$ unconventional gap with $\Delta ({\bf k}) \propto\cos k_x +\cos k_y $ and a realistic amplitude ${\mathrm{max}}_{\mathbf{k}}\,\Delta(\mathbf{k})\simeq4.2\,\mathrm{meV}$ at $T=15$\,K. Lastly, for CeCoIn$_5$, shown  in Fig.\,\ref{symmetry}(i), we find a higher-harmonic $d_{x^2-y^2}$-gap symmetry with $\Delta ({\bf k}) \propto\cos 2k_x -\cos 2k_y $ and $\mathrm{max}_{\mathbf{k}} \,\Delta(\mathbf{k})\simeq0.87\,\mathrm{meV}$ at $T=2$\,K which is also in reasonable agreement with experiment \cite{Fasano2018}.

%\textbf{Analysis of unconventional symmetry.}
Our calculations reveal that {the strongly momentum-dependent}
gap symmetries in these superconductors can be 
obtained solely via isotropic EPI provided that vertex corrections to the electron self-energy are taken into account. 
A crucial ingredient for this result is the fact that the vertex correction in Eq.\,(\ref{vertex}) is inherently \textit{momentum-dependent} regardless of the bare EPI being momentum-independent. To better {understand} this point, we calculate the static vertex by plugging our selfconsistent results back into Eq.\,(\ref{vertex}) and taking $m=m'=0$. In the static case all channels share the same vertex which is then a scalar given by
\begin{eqnarray}\label{vertex0}
\Gamma\ph_{{\bf q},0}&=&1 + g^2_0 T\sum_{m''} D_{m''}\frac{1}{2}\rm{Tr}\left\{\hat{\Lambda}\ph_{{\bf q},m''}\right\} .
\end{eqnarray}
The calculated static vertices, shown in Figs.\,\ref{symmetry}(b), (f), and (j), have momentum structures {that share}  similar characteristics. $\Gamma\ph_{{\bf q},0}$ is strongly peaked and negative, i.e., repulsive, for wavevectors that almost or perfectly coincide with the nesting wavevectors that correspond to the susceptibility peaks of Figs.\,\ref{symmetry}(d),(h), and (l) and positive otherwise. These large repulsive peaks allow pairing with a sign-change across the BZ thus providing a mechanism for unconventional superconductivity in a manner similar to spin fluctuations. However,  in contrast to the latter, where the small wavevector part of the repulsive interaction can interfere destructively to pairing \cite{Schrodi2020_3,Yamase2020,SchrodiMultiChan}, in our case the EPI remains attractive at smaller wavevectors and therefore can contribute to the unconventional pairing. %Notably, due to the fact that the resulting gap is unconventional, the direct Coulomb repulsion, had we included it in the calculations, would have been completely avoided in this vertex corrected EPI mechanism. 
We {therefore} conclude that the vertex-corrected EPI in all three prototypical examples 
is selfconsistently optimized so as to maximize {the} pairing. Mathematically, this is possible {because
the selfconsistent procedure corresponds to minimizing the system's free energy, and}
the superconducting gap and vertex functions are allowed to {mutually influence}
each other during the selfconsistent cycle. Physically, this can be understood as the result of the formed Cooper pairs dynamically renormalizing the EPI and vice versa.

To take our analysis further, we draw a much more simplified picture where we make a one-loop approximation to Eq.\,(\ref{vertex0}) by using noninteracting Green's functions, $\hat{G}^{(0)}_{{\bf k},n,m}=[i\omega_m-\xi_{{\bf k},n}\hat{\rho}_3]^{-1}$.
The resulting bare static vertices, $\Gamma^{(0)}_{{\bf q},0}$, calculated for our three cases are shown in Figs.\,\ref{symmetry}(c), (g), and (k). For cuprate and Fe-based systems, these are roughly similar to their respective interacting vertices [see Figs.\,\ref{symmetry}(b),(f)]. However for CeCoIn$_5$, that has a more complex FS, the bare vertex shown in Fig.\,\ref{symmetry}(k) is positive everywhere in the BZ in contrast to the respective selfconsistent interacting vertex of Fig.\,\ref{symmetry}(j). This case presents a clear example where inclusion of the back-reaction of the superconducting gap to the vertex function is essential for finding the correct gap symmetry and therefore for the occurrence of unconventional pairing. This also {implies} that although the much-easier-to-calculate bare vertex function may serve as an indicator of possible unconventional superconductivity due to vertex-corrected EPI, it may nevertheless miss cases with more complex FSs, {and} therefore, full selfconsistent calculations 
are indispensable for accurate  predictions. 

As can be seen in Fig.\,\ref{symmetry}, apart from possible different signs, the bare vertices are very similar with the respective bare susceptibilities for each system. 
%This also holds in the interacting case (see \cite{supplement}). 
In fact, the static charge susceptibility can be calculated from Eq.\,(\ref{suscl}) as ${\cal X}_{{\bf q},0}=(1/2)T\sum_{m''}\rm{Tr}\bigl\{\hat{\Lambda}\ph_{{\bf q},m''}\bigl\}$.
The results shown in Figs.\,\ref{symmetry}(d), (h), and (l) correspond to the noninteracting case, 
${\cal X}^{(0)}_{{\bf q},0}$.
One can now observe that the second term on the right of Eq.\,(\ref{vertex0}) is proportional to the charge susceptibility weightened by the phonon propagator $D_m$. Since in our simple Einstein phonon picture this weightening takes place only in frequency space, it is responsible for making parts of the EPI repulsive and parts of it attractive, while the overall momentum structure follows that of the charge susceptibility.

\begin{figure}[t!]
	\centering
	\includegraphics[width=\linewidth]{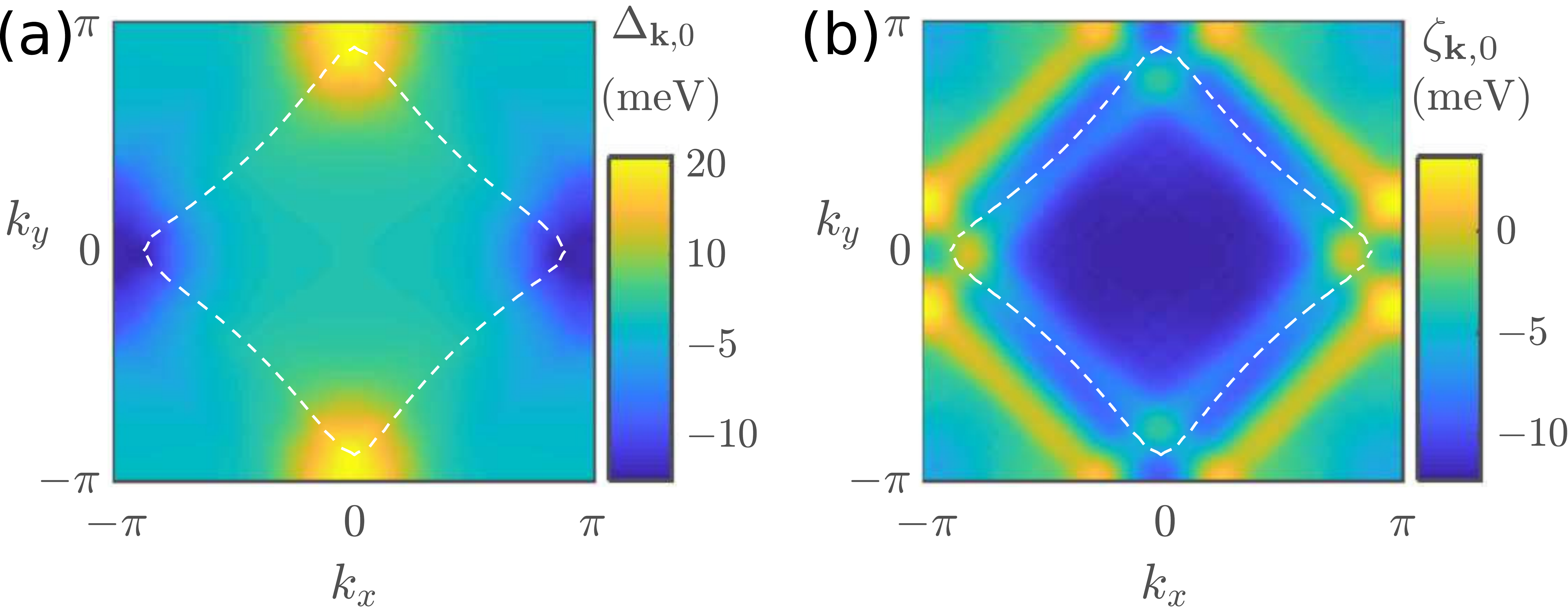}
	\vspace*{-0.6cm}
	\caption{{Selfconsistently calculated nematic superconductivity.} (a) The calculated nematic superconducting  {gap  $\Delta_{\mathbf{k},0}$ for {a} cuprate Fermi} surface {(white dashed line)}
	in the slightly hole-doped regime, $(C,t,t',\mu)=(0,-1/4,1/10,$ $-0.11)\,\mathrm{eV}$. (b) 
	The computed zero-energy chemical potential renormalization, $\zeta_{\mathbf{k},m=0}$. }
	\label{nematicity}
\end{figure}

%\textbf{Nematic superconductivity.}
Another interesting 
{outcome} of our theory {for} the cuprates is the possibility to obtain selfconsistently nematic superconductivity. The so-called nematic state, in which $C_4$ rotational symmetry is 
{reduced} to a $C_2$ orthorombic structure, occupies a well-established region in the temperature-doping phase diagrams of cuprate superconductors \,\cite{Fradkin2010}. 
%There is experimental and theoretical support that nematicity does not compete but coexist with superconductivity\,\cite{Vojta2009,Su2011}.
 For convenience of the calculations, we here use  $g_0=170$\,meV \cite{noteMats}. 
% \footnote{Using a larger $g_0$ leads to a higher $T_c$, therefore allowing us to work at higher temperatures that demand fewer Matsubara frequencies and are thus less computationally demanding. For $g_0 = 170$ meV and the same parameter set as used in Fig.\ 1(a) we find the same $d$-wave symmetry but with larger gap value and $T_c \approx 140$\,K.}
 We also shift the dispersion towards hole-doping by setting $\mu=-0.11\,\mathrm{eV}$. The resulting FS is drawn in Fig.\,\ref{nematicity} (white-dashed lines). The rest of the parameters are unchanged. 
% Thus all input to our calculation is per default tetragonal. Nevertheless, when s
 Solving the vertex-corrected Eliashberg equations 
 we observe a spontaneous change from $C_4$ to $C_2$ BZ symmetry in all three functions $\Delta_{\mathbf{k},m=0}$, $\zeta_{\mathbf{k},m=0}$, and $Z_{\mathbf{k},m=0}$. The first two are shown in Fig.\,\ref{nematicity}(a) and (b), respectively.
 
The superconducting gap of Fig.\,\ref{nematicity}(a) is an admixture of $d_{x^2-y^2}+s$ symmetries that lowers $C_4$ symmetry to $C_2$ and is therefore nematic. Inspection of the chemical potential renormalization in Fig.\,\ref{nematicity}(b) similarly reveals a two-fold symmetric energy band renormalization which we can identify as an induced Pomeranchuk order parameter \cite{Livanas2015}. Nematicity appears only at low energies, we always {obtain} anisotropic $s$-wave symmetry solutions for $m\gg1$. It disappears for $T>T_c$ where we find a restored $C_4$ symmetry in $\zeta_{\mathbf{k},0}$ and $Z_{\mathbf{k},0}$. These results indicate that the vertex-corrected renormalized EPI may very well be simultaneously pairing even in seemingly competing pairing channels, such as those of $s$-wave and higher angular momentum, which, depending on the underlying bandstructure, may lead to nematic superconductivity being the energetically {favored} solution.

%\textbf{Discussion.}
Apart from the unconventional {gap} symmetry,
another  
{frequent} argument against the relevance of the EPI in unconventional high-$T_c$ superconductors is a seemingly small coupling constant\,\cite{Giustino2008}. {However,} full-bandwidth Eliashberg theory generally includes Cooper pairing {of states} away from the Fermi level which may contribute to the gap size and $T_c$\,\cite{Aperis2018}. A similar behavior {can} be expected in 
vertex-corrected full-bandwidth theory \cite{Schrodi2020}. Focusing on the cuprate case, we carried out the computationally heavy task of solving for the complete selfconsistent temperature dependence of our vertex-corrected Eliashberg equations. % \cite{supplement}.
Within our chosen parameter set, we find a realistic $T_c\approx 52$\,K. We 
{furthermore}  estimated that for $T>T_c$, the effective electron-phonon coupling is $\lambda_m\approx0.34$ {(see SM \cite{supplement}).}
% \cite{supplement}).}
%Remarkably, we find $\lambda_m\approx0.34$.
 Therefore, seemingly weak coupling values are  {nonetheless} compatible with phonon-mediated high-$T_c$ unconventional superconductivity.

We note that our findings do not exclude that spin fluctuations contribute to Cooper pairing, since both EPI and spin fluctuations can lead to a pairing symmetry derived from the FS nesting, and hence, both mechanisms can act cooperatively. To establish unambiguously the relative size of their contributions, selfconsistent simulations within  multichannel vertex-corrected  Eliashberg theory will be required, to treat both bosonic mediators on equal footing. 
Lastly, we mention that we have neglected the influence of the Coulomb repulsion, which does not affect our results, as the direct electron-electron Coulomb repulsion cancels out for an unconventional order parameter. 

{Our vertex-corrected Eliashberg-theory calculations} provide proof of principle that isotropic 
EPI can give rise to unconventional superconductivity.
{This phonon-mediated unconventional superconductivity relies on Fermi surface nesting and moderate  nonadiabaticity that are present in many unconventional superconductors \cite{Uemura1989,Scalapino2012}. The latter gives rise to a repulsive EPI which is anti-pairing in the $s$-wave channel \cite{Schrodi2020} while the former makes this interaction peak near the nesting  wavevector(s) and thus favorable for unconventional pairing in a similar fashion as purely electronic mechanisms like spin fluctuations. 
Our results {show that the conventional EPI can mediate unconventional superconductivity and highlight}
the possibility that electron-phonon coupling plays a substantial role in any superconductor with unconventional gap symmetry including the high-$T_c$s.}

%Our findings call for a critical re-examination of the accumulated results regarding the mechanisms of superconductivity and establish the EPI on the list of primary mechanisms for unconventional Cooper pairing with broad applications in the field of superconductivity. {\green discuss}

%\textbf{Acknowledgements.}
This work has been supported by the Swedish
Research Council (VR), the R\"ontgen-Angstr\"om Cluster,
and the Knut and Alice Wallenberg Foundation (Grant No.\ 2015.0060). The calculations were enabled by resources provided by the Swedish National Infrastructure for Computing (SNIC) at NSC Link\"oping, partially funded by the Swedish Research Council through grant agreement No.\ 2018-05973.\\[-0.2cm]

%\bibliographystyle{apsrev4-1}
%\bibliography{dwaveBib.bib}{}

\begin{thebibliography}{43}
\bibitem [{not()}]{noteMats}%
  \BibitemOpen
  \href@noop {} {}\bibinfo {note} {Using a larger $g_0$ leads to a higher
  $T_c$, therefore allowing us to work at higher temperatures that demand fewer
  Matsubara frequencies and are thus less computationally demanding. For
  $g_0=170$\,meV and the same parameter set as used in Fig.\,\ref{symmetry}(a)
  we find the same $d$-wave symmetry but with larger gap value and
  $T_c\approx140$\,K.}\BibitemShut {Stop}%
\end{thebibliography}

%

\end{document}